\documentclass[aip,reprint]{revtex4-2}

\draft % marks overfull lines with a black rule on the right

\usepackage{graphicx}% Include figure files
\usepackage{dcolumn}% Align table columns on decimal point
\usepackage{bm}% bold math
\usepackage{amsmath}
\def\be{\begin{equation}}
\def\ee{\end{equation}}
\def\bea{\begin{eqnarray}}
\def\eea{\end{eqnarray}}

\begin{document}

\title{Dynamics of the Surface Growth Resulted from Sedimentation of Spheres in a Hele-Shaw Cell Containing a Low-Viscosity Fluid}% Force line breaks with \\

\author{Vahideh Sardari}
\email{v.sardari@iasbs.ac.ir}
\affiliation{Department of Physics, Institute for Advanced Studies in
Basic Sciences (IASBS), Zanjan 45137-66731, Iran}

\author{Fatemeh Safari}
\email{fatemeh.safari@iasbs.ac.ir}
\affiliation{Department of Physics, Institute for Advanced Studies in
Basic Sciences (IASBS), Zanjan 45137-66731, Iran}

\author{Maniya Maleki}
\email{m\_maleki@iasbs.ac.ir}
\affiliation{Department of Physics, Institute for Advanced Studies in
Basic Sciences (IASBS), Zanjan 45137-66731, Iran}

\date{\today}% It is always \today, today,

\begin{abstract}

In this paper, we investigate the dynamics of surface growth resulting from sedimentation of spherical granular particles in a fluid environment, using experiments and simulations. In the experimental part, spherical polystyrene particles are poured down from the top of a vertical Hele-Shaw cell and form a 1+1-dimensional growing surface. The surface roughness is obtained from the images and the growth and roughness exponents are measured. In the numerical simulation part, the surface growth process is simulated using the Molecular Dynamics method, considering the interactions between the grains; and the exponents are calculated. In this method, unlike conventional simulation models, instead of a discrete deposition law, the dynamics of the individual particles throughout the process are obtained considering  different forces acting on the particles. Finally, the simulation results are compared with the experiment, and we see a very good agreement between them. We find different values for the exponents using different methods, which indicates that the system is multi-affine and does not obey scaling laws of affine models.
\end{abstract}

\maketitle

%\tableofcontents

\section{Introduction}

The time evolution of surfaces and interfaces is a common phenomenon in experiments, in nature, and even in our daily lives. This phenomenon can be seen in all scales from the microscopic to the macroscopic. As an example, at the earth's surface, factors such as erosion and weathering cause temporal changes in its surface. Other examples include  phenomena such as liquid flow in porous media, fire tongues, wetting fronts, bacterial colony growth, and tumors, as well as deposition processes ranging from sedimentation in geology to electrochemical deposition. Surface morphology can affect the physical and functional performance of industrial processes. For example, the roughness affects the optical properties of thin films, the adhesion of layers to each other, their friction, or the electrical properties of layers. Therefore, the study of surface evolution in the growth phenomena can be of great help in understanding and controlling this phenomenon, for technological applications \cite{Barabasi, film}. 

Two general approaches are commonly used to theoretically investigate surface growth issues. The first approach involves discrete growth models, which are usually based on computer simulations, and the second approach is to study the dynamics of surfaces using stochastic differential equations. In this approach, micro-scale details are neglected and all system details are summarized in a limited number of parameters, which is useful for studying the asymptotic scaling behavior of the system. In the past few decades, several discrete and continuum growth models have been proposed to describe growth dynamics that well describe the characteristics of various growth processes \cite{Family3}.
These discrete models include Random Deposition (RD) model, Random Deposition with Surface Diffusion \cite{Family2}, Ballistic Deposition (BD) model, as well as nonlinear discrete models such as the Eden Growth Model (ED) \cite{Eden}, the Solid-on-Solid model (SOS) \cite{BD}, and the Restricted Solid-on-Solid model (RSOS) \cite{RSOS}. The simulation results of the discrete models are very similar to some real phenomena. For example, the ballistic deposition model gives a good description of the real paper burning process, and the uneven surface formed in this way \cite{burn}. 

In addition to discrete models, continuum equations such as the Edward-Wilkinson (EW) \cite{EW} equation and the Kardar-Parisi-Zhang equation (KPZ) \cite{KPZ} also provide a suitable description for real surface growth processes. It can be said that the KPZ equation is the simplest possible equation of motion to describe the dynamics of an intersection that encompasses all the non-obvious growth behaviors such as irreversibility, nonlinearity, randomness, and localization \cite{KPZ}. Due to the complexity of the surface growth experiments, it is difficult to find experimental systems that are well-described by the theoretical models or simulations. Such that only in a few experiments, including the colony growth of Bacillus Subtilis \cite{Wakita}, as well as eukaryotic Vero cell \cite{coloni}, slow-burning of the paper \cite{burn}, the deposition of the elliptical particles on the edge of an evaporating colloidal droplet \cite{Yunker, Yunker2}, and interfaces of topological-defect turbulence in the electroconvection of nematic liquid crystals \cite{Sano} the KPZ exponents have been reported. 

In this paper, we have studied the deposition of polystyrene sedimenting particles in ethanol in a quasi-two dimensional system, resulting in a 1+1 dimensional growing surface. We have done both experiments and simulations. In the simulation section, we have used the MD molecular dynamics method instead of the usual methods, considering all the interactions and details of particle shedding. Unlike previous models, our model is not latticed and after deposition, particles can still be rearranged under the influence of forces exerted by other particles. The effect of upper layers pressure, friction, and inelastic collisions are all considered. In the experimental section, we have performed controlled and precise experiments, which finally show a good agreement with the simulation results. We have calculated the growth and roughness exponents with different methods and showed that the system is multi-affine and does not obey conventional scaling laws. 

\section{Scaling concepts}

Scaling, based on the general concepts of the scale-invariance and fractal behavior, is a standard tool in the study of growing surfaces and is used to study a variety of theoretical models and experiments. Scaling analysis identifies and categorizes growing surfaces with a set of critical exponents. These exponents determine the morphology and dynamics of growing surfaces. Usually, in order to describe the surface growth quantitatively, two functions of surface mean height $\bar{h}$ and roughness width $w$ (the standard deviation of height) are calculated first, which are defined as \cite{Barabasi, Family1}
\bea
&\bar{h}(t)=\frac{1}{L}\sum_{i=1}^L h(i,t),\\
&w^2(L,t)=\frac{1}{L}\sum_{i=1}^{L} \left[h(i,t)-\bar{h}(t)\right]^2,
\eea
where $h(i, t)$ indicate the height of the i-th column at time $t$ and $L$ is the size of the system. If the deposition rate is constant, the mean height increases linearly with time $\bar{h}\sim t$. A typical form of the evolution of the roughness width with respect to time has two distinct regions that are separated by a crossover time $t_x$ . Initially, the roughness width increases with time and satisfies the following relation \cite{Family1, Family2}
\begin{equation}
w(L,t)\sim t^{\beta}\  \  \  \  \  \  (t\ll t_x),
\end{equation}
where exponent $\beta$  is called "the growth exponent" and determines the time-dependent dynamics of the growth process. The increase of the roughness width does not continue indefinitely, but reaches a saturation value that is represented by $w_{sat}$. If the saturated roughness width is plotted for different values of $L$, it is observed that with increasing the size of the system size $L$, the saturation width also increases as
\begin{equation}
w_{sat}(L)\sim L^{\alpha} \  \  \  \  \  \  (t\gg t_x),
\end{equation}
and exponent $\alpha$ is called "the roughness exponent", which characterizes the size-dependent behavior of the saturated roughness. The above relation is often expressed by the following relation, which is known as the Family-Vicsek scaling relation \cite{Family1, Family2}
\bea
w(L,t)\sim L^{\alpha} f\left(\frac{t}{L^z} \right), 
\eea
where $f$ is called the scaling function and shows the following asymptotic behavior
\begin{align}
f(u) & \sim u^{\beta}\  \  \  \  \  \  \  \  \  \  &(u \ll 1),\\
f(u) &=const. \  \  \  \  \  \  \  \  \  &(u\gg 1), 
\end{align}
and $z=\alpha / \beta$.

\section{Experimental Procedure}

\begin{figure}[h!]%[htbp] 
\centering
\centerline{\includegraphics[width=0.96 \columnwidth]{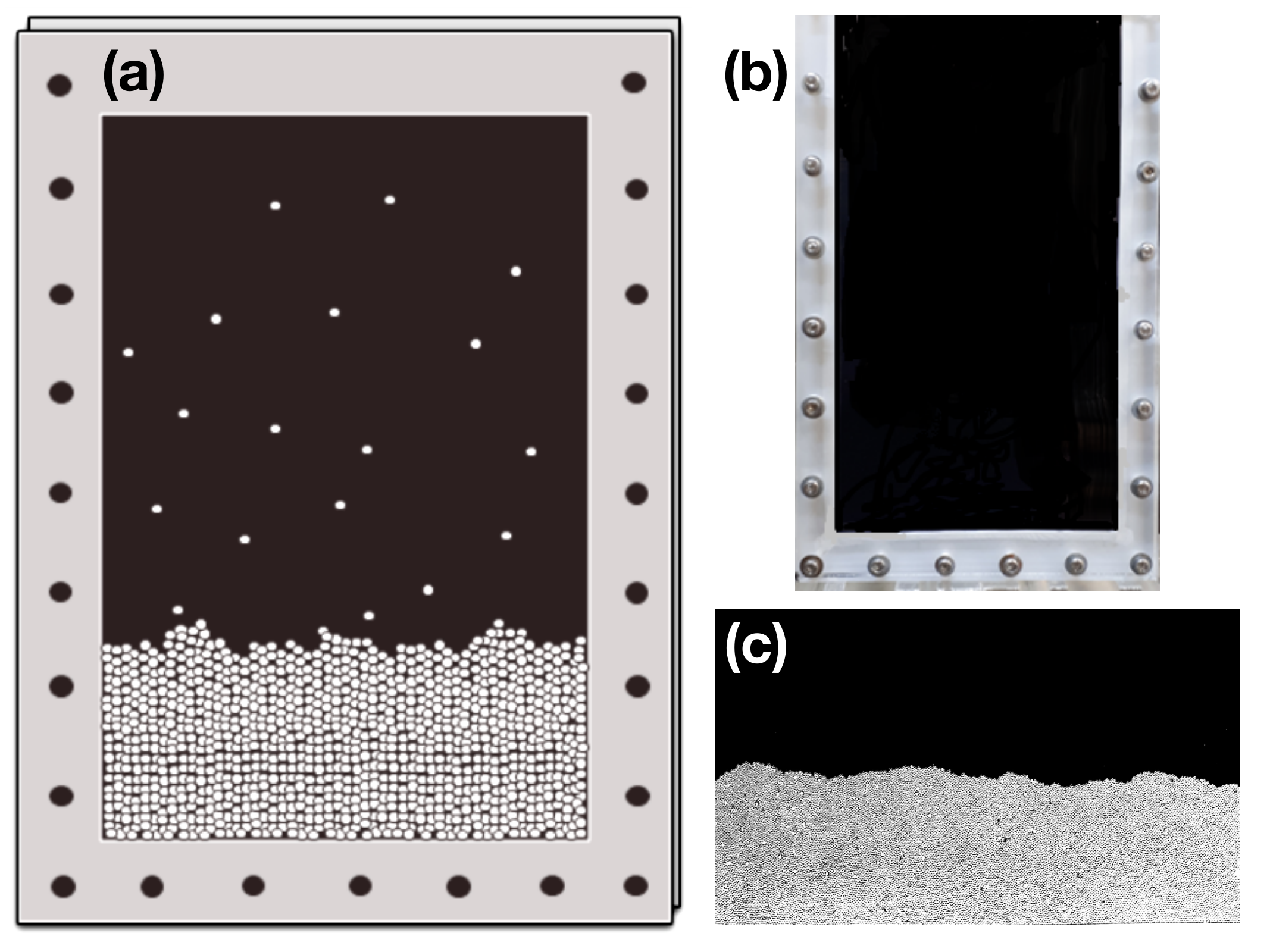}}
\caption{The experimental setup: (a) schematic: Hele-Shaw cell consists of two Plexiglas plates, with a distance a little more than a particle size. The cell is filled with Ethyl Alcohol and Polystyrene particles are poured and settled to the bottom of the cell and build a rough surface; (b) the cell used in experiments; (c) a snapshot of the formed growing surface in experiment.}
\label{setup}
\end{figure}

\begin{figure}[h!]%[htbp] 
\centering
\centerline{\includegraphics[width=0.9 \columnwidth]{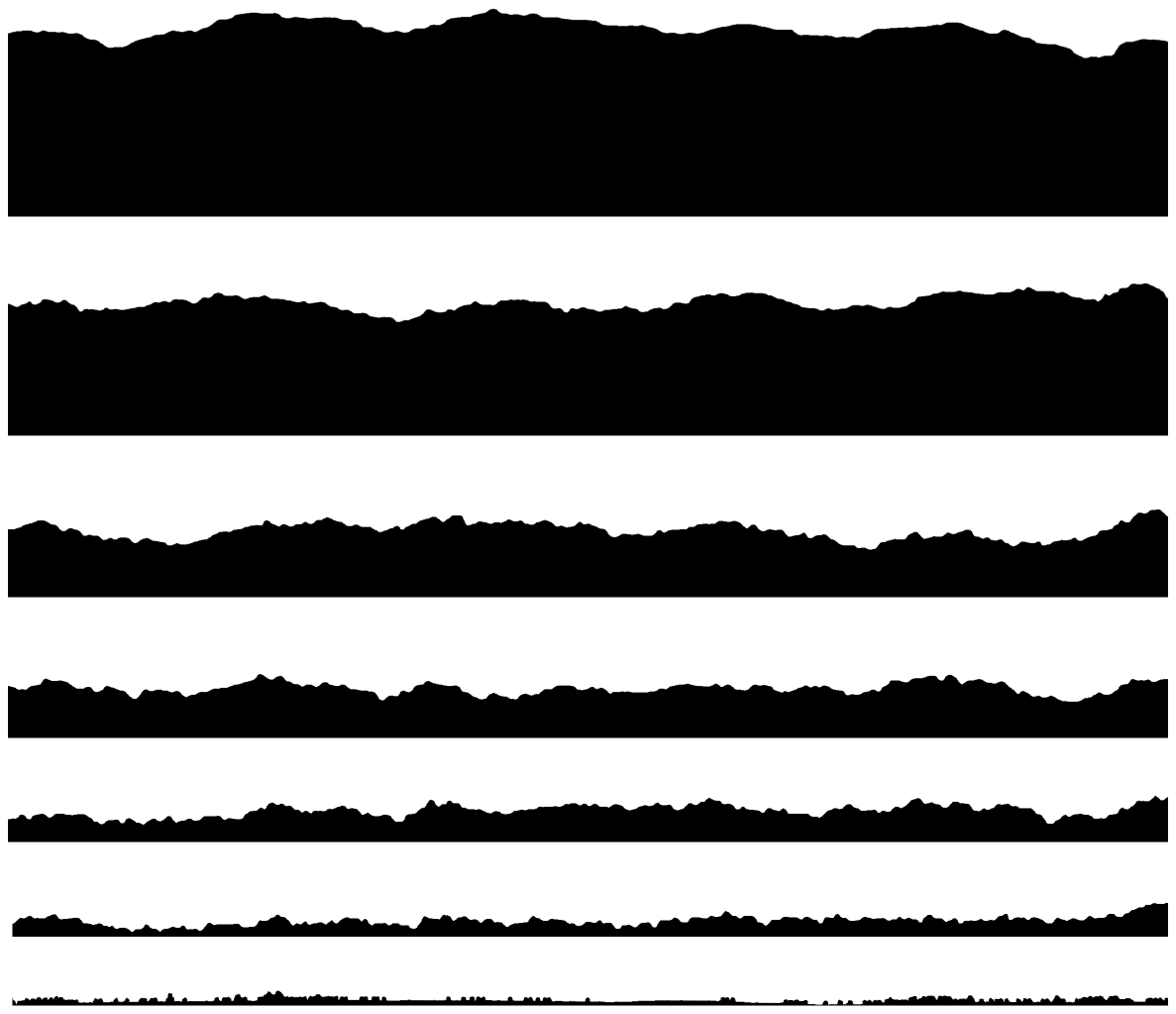}}
\caption{Snapshots of the surface growth in experiments. The time and average height increase from bottom to top.}
\label{growth-exp}
\end{figure}

In experiments, we use a vertical Hele-Shaw cell which consists of two transparent plexiglass plates with dimensions of 25 cm $\times$ 40 cm separated by a strip of teflon of a thickness a little more than the diameter of the beads (600$\pm$50 $\mu$m). In addition to sealing the test chamber, the purpose of this gasket is to create the desired distance between the two plexiglas plates. The distance between the plates, is obtained by measuring the volume of fluid that completely fills the cell.The plexiglas plates used are transparent and essentially rigid, so the distance between the plates can be considered constant at different points. As a result, during the deposition of particles, a growing front is formed which behaves close to a 1 + 1-dimensional  surface, due to the particle size and the distance between the plates. We use spherical polystyrene particles with a diameter of 500 microns, made by Sigma-Aldrich company. To perform the experiments, we fill the space between the plates up to a height of 30 cm with 96\% Ethanol, and the polystyrene particles are poured into the cell randomly, uniformly, and in small numbers, through the open head of the cell (Fig. \ref{setup}). After the particles settle to the bottom of the cell and build a rough surface, at each stage, we take photos of the formed surface. Finally, we use ImageJ software to get binary images, then using a MATLAB code, the experimental interface is tracked and we obtain the roughness and growth exponents. Fig. \ref{growth-exp} shows the stages of surface growth in the experiment.

\section{Simulation}

To simulate the surface growth process, we used the LAMMPS Molecular Dynamics Simulator. Since our physical system contained granular material, we used the LAMMPS granular package, which included Hertzian interactions between particles, as well as between particles and the wall. This package uses the following formulas for the normal and tangential forces ($F_n$ and $F_t$) between two granular particles $i,j$ of radii $R_i, R_j$. When the distance $r_{ij}$ between two particles is less than $d=R_i+R_j$ (if $r>d$, there will be no force between the particles) \cite{LAMMPS}
\bea
F_{n_{i,j}}=\sqrt{\frac{\delta_{i,j}}{d}}\left( k_n \delta_{i,j} n_{i,j}-\gamma_n m_{eff}v_{n_{i,j}}\right),\\
F_{t_{i,j}}=\sqrt{\frac{\delta_{i,j}}{d}}\left(-k_t \delta_{i,j} t_{i,j}-\gamma_t m_{eff}v_{t_{i,j}}\right).
\eea
In the above equations, $\delta_{i,j} = R_i + R_j - r_{ij}$, and $k(n,t)$, $\gamma(n,t)$ and $m_{eff}$ are the elastic constant, viscoelastic damping constant and the effective mass of the particles, respectively. $v_{n_{i,j}}$ and $v_{t_{i,j}}$ represent the normal and tangential components of the relative velocity of particles $i$ and $j$. The dimensions of the simulation box was chosen to be the same as the experiments, and we randomly scattered a certain number of particles (approximately 100 particles at each stage) at a given and fixed height from the bottom of the box. The particles moved to the bottom of the container due to the effective gravity force and the 1 + 1 dimensional surface began to grow. Because the particles are in a fluid environment, in addition to the gravity force, Archimedes’ force and viscous drag force affect the particles. Therefore, in the simulation, we used an effective acceleration that included both gravity and Archimedes forces
\be
g_{eff}=g( \rho_{p}-\rho_s)/\rho_p,
\ee
where $g$, $\rho_p$ and $\rho_s$ are gravity acceleration, particle density, and fluid density, respectively. For the viscous drag force on a particle moving in a fluid, we can use Stokes’ law in the low Reynolds number regime. It is important to note that the effective viscosity of the fluid that the particles feel in our system, is more than the amount for a particle in an unbounded fluid (approximately 3.5 times in our experiment) when the fluid is constrained between two walls with a short distance \cite{Swan}, thus we use the effective viscosity in the drag force. 

Once all the deposited particles at each stage are fixed in their place and a new layer of 1 + 1 dimensional surface is formed, the next series of particles is poured into the box. The simulation continues until all the particles are deposited and the surface roughness is saturated. Like the experimental part, we use ImageJ software to get binary images, then use a MATLAB code to obtain the boundary of the sediment layer.  Then, we obtain the roughness and growth exponents. In Figure 3, a snapshot of the growing surface and its transformed binary picture are presented.

\begin{figure}[h!]%[htbp] 
\centering
\centerline{\includegraphics[width=0.96 \columnwidth]{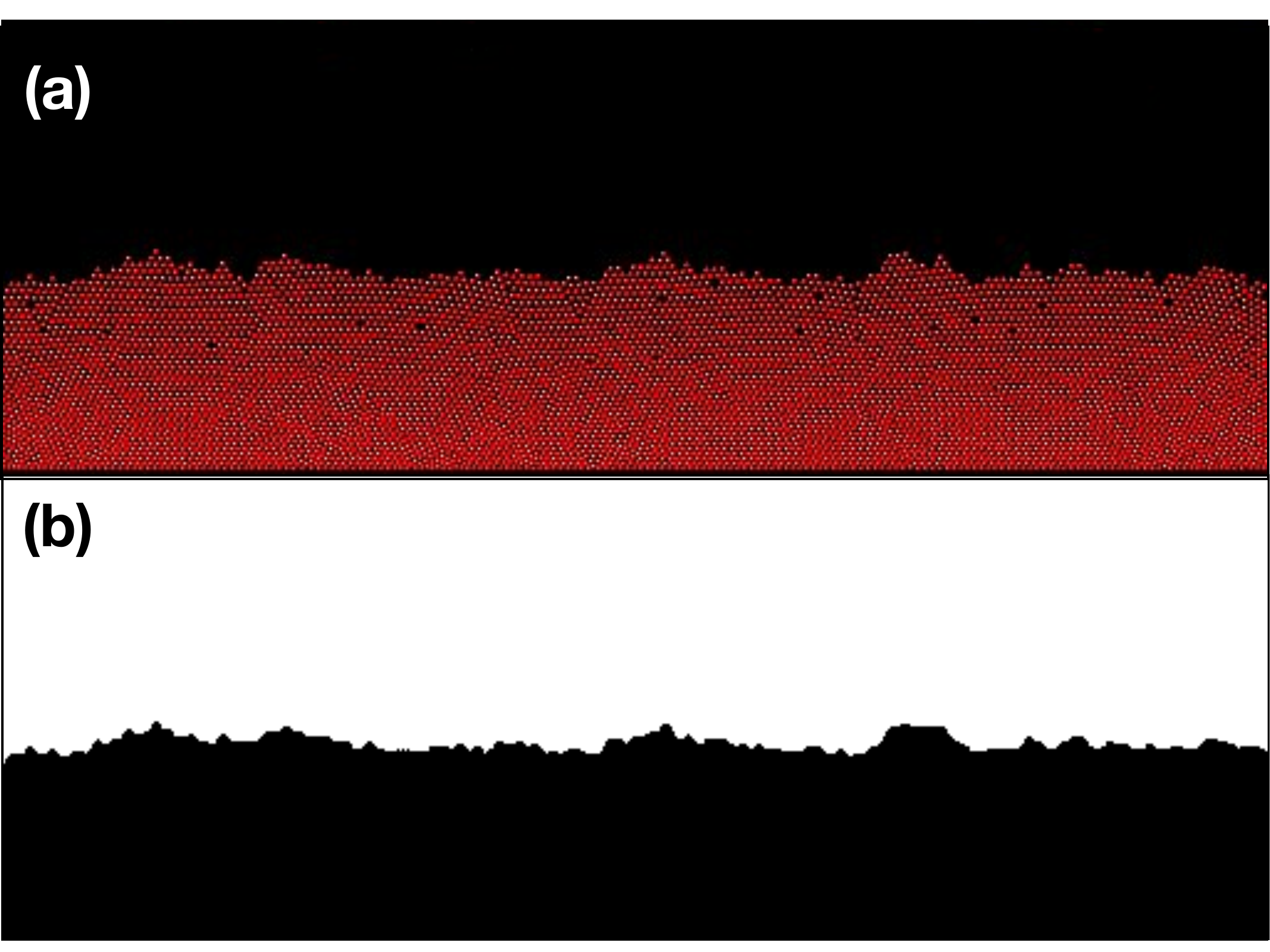}}
\caption{A snapshot of the surface formed in the simulation (a) and its transformed binary picture (b).}
\label{sim}
\end{figure}

\section{Results}

\begin{figure}[h!]%[htbp] 
\centering
\centerline{\includegraphics[width=0.96 \columnwidth]{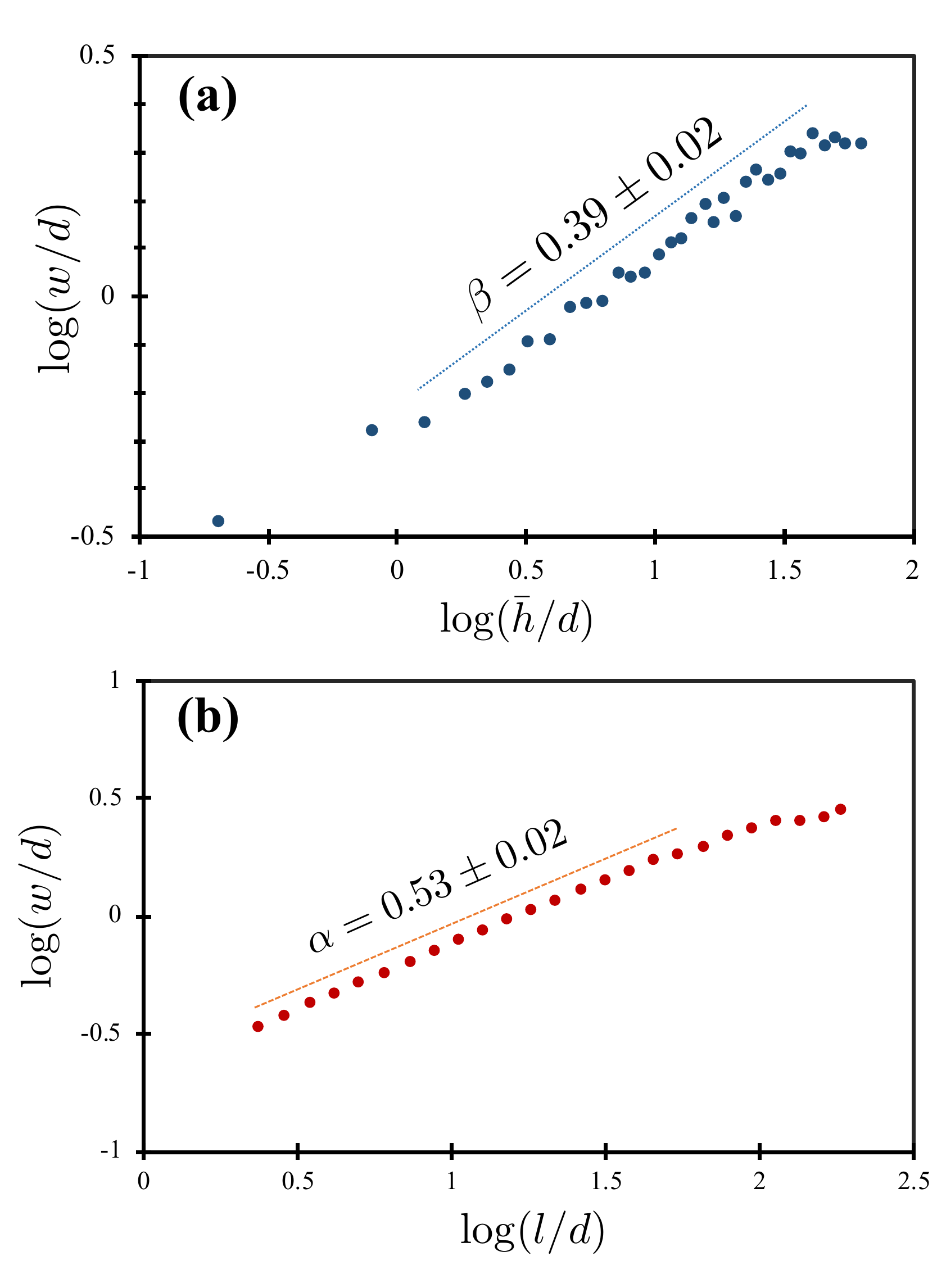}}
\caption{Experimental results: (a) the growth exponent and (b) the roughness exponent.}
\label{exp-alphabeta}
\end{figure}

\begin{figure}[h!]%[htbp] 
\centering
\centerline{\includegraphics[width=0.96 \columnwidth]{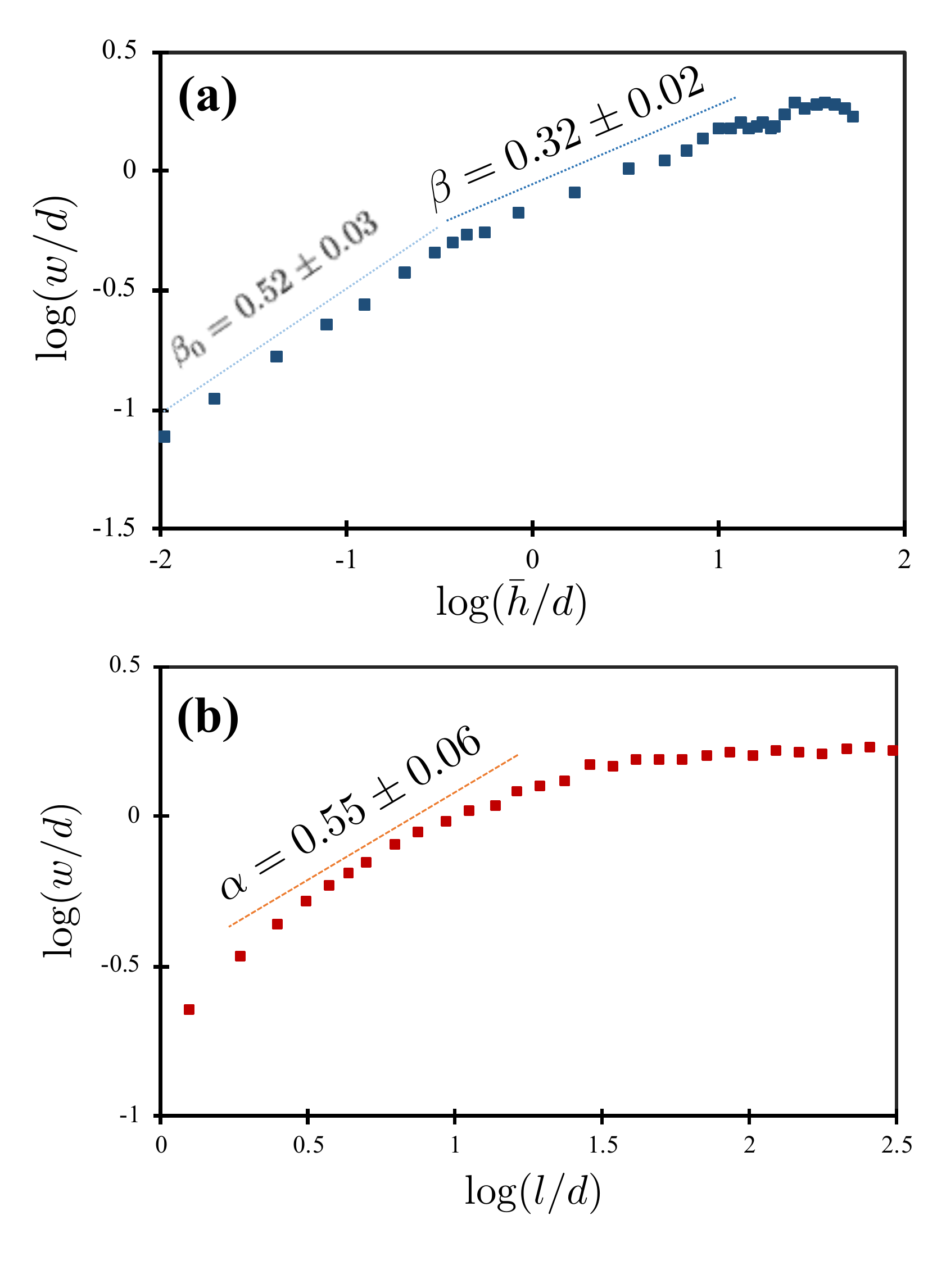}}
\caption{Results of simulation: (a) the growth exponent and (b) the roughness exponent.}
\label{sim-alphabeta}
\end{figure}

As mentioned earlier, for many growth systems, the width of the interface $w(t)$ increases as a power of time, and then saturates at a value that increases as a power law of the system size. These exponents do not show how rough the surface is, but they give us a good estimate of how the roughness changes with time and the system size. One of the most common methods for determining roughness and growth exponents is to use Equations (3) and (4). To obtain the growth exponent, we need to plot the width of the deposition boundary over time in a log-log scale. The slope of this plot (before reaching saturation) gives us the growth exponent $\beta$. For this purpose, we must obtain the image of all the stages of particle deposition at different times, after the complete stability of the particles in their place. Since in our experiment, the rate of shedding of particles is not constant in time, we use average height instead of time in Equation (4) $w\sim \bar{h}^\beta$. The results of experiments and simulations are shown in Figures \ref{exp-alphabeta}(a) and \ref{sim-alphabeta}(a), where the logarithm of roughness $w$ is plotted versus  the logarithm of mean height $\bar{h}$ (both are first scaled to the particle diameter $d$).

 In order to determine the roughness exponent using Equation (4), it is necessary to repeat the experiment at different sizes of the system $L$. It is very difficult to provide these conditions in the experiment. To solve this problem, instead of the total width of the surface, we calculate the local width of the saturation boundary (Equation (11)). In this equation, $\langle \rangle$ indicates the ensemble averaging. To do the averaging, we divide the surface into windows of length $l$, and average the results obtained from all the windows. Finally, the local width will be related to the length of the window in the following form 
\be
w^2(l,t)=\left< \frac{1}{l} \sum_{x=x_0}^{x_0+l}\left[h(x,t)-\bar{h}(t)\right]^2 \right>_{x_0},
\ee
where $\bar{h}(t)$ is the average height in the selected window of observation. Using Equation (4) for the local width $w \sim l^\alpha$, exponent $\alpha$ is obtained, as seen in Figures \ref{exp-alphabeta}(b) and \ref{sim-alphabeta}(b).

 The growth and roughness exponents obtained from the experiments are $\beta= 0.39 \pm 0.02$ and $\alpha = 0.53 \pm 0.02$ (Figure \ref{exp-alphabeta}). The roughness versus height plot (in log-log scale) obtained from the simulations, exhibits two different slopes: for  $h/d<0.3$ (meaning before having a complete monolayer on the bottom) an initial slope of  $\beta_0= 0.52 \pm 0.03$ and afterwards, a second slope $\beta= 0.32 \pm 0.02$ is observed. The roughness versus system length $l$ plot resulted from the simulations (in log-log scale) gives a mean slope of  $\alpha = 0.55 \pm 0.06$. There is a good agreement between the results of the experiments and the simulations, which was not observed by others in previous works. These values obtained for the exponent are close to the theoretical predictions based on the Kardar, Parisi, and Zhang equation and Ballistic deposition in $1+1$ dimensions.

\begin{figure}[h!]%[htbp] 
\centering
\centerline{\includegraphics[width=0.95 \columnwidth]{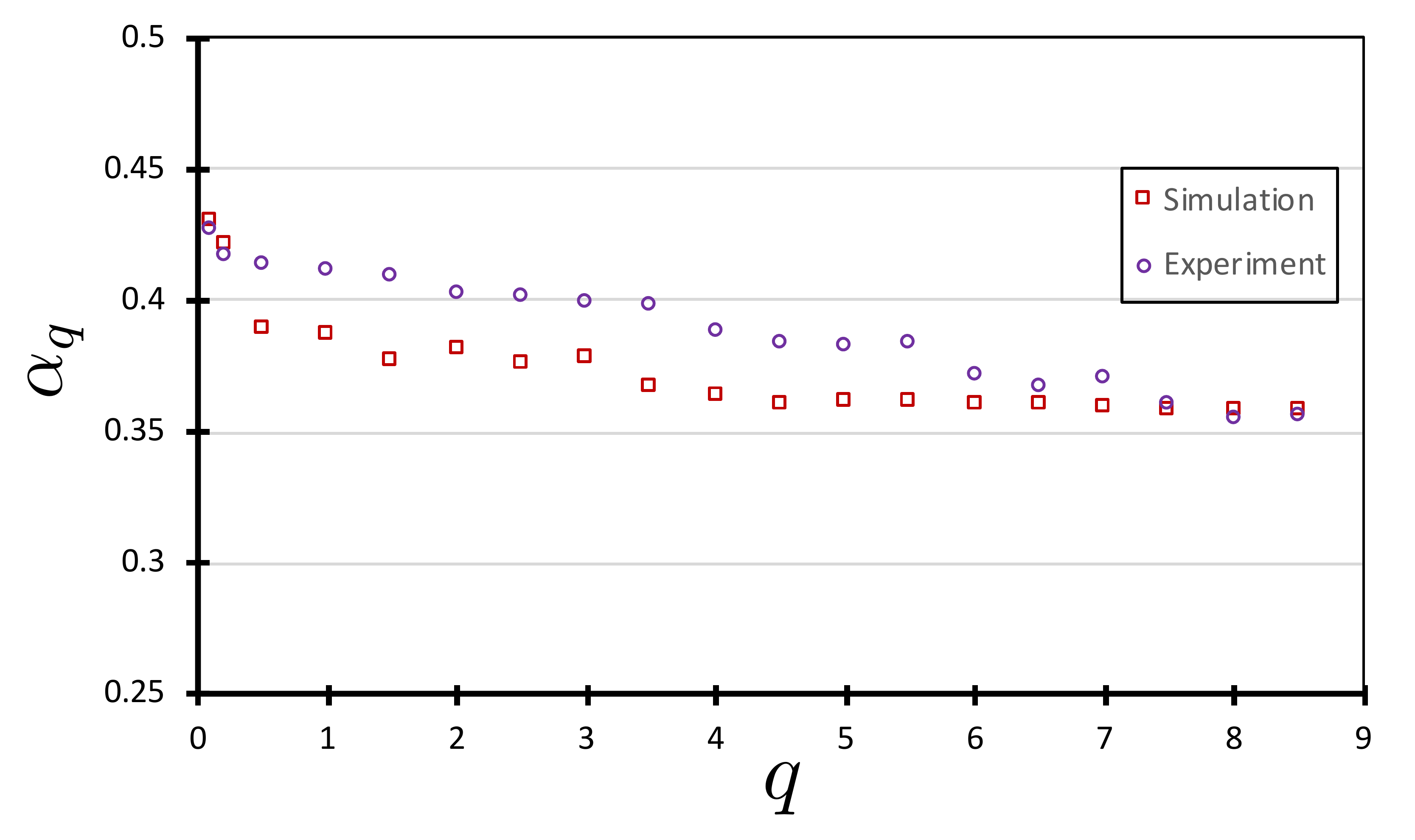}}
\caption{The $q$'th order of the roughness exponent derived with spatial correlation function.}
\label{alpha-q}
\end{figure}

Another way to calculate the roughness exponent is using the spatial correlation function \cite{Barabasi, multi}. This method also can be used to check the multifractality of the surface. The $q$th order spatial correlation function $C_q$ is defined as
\be
C_q(l)=\left[ \left< \left(h(x,t)-h(x+l,t) \right)^q \right>_{x,t} \right]^{1/q}\sim l^{\alpha_q},
\ee
where the distance $l$ is smaller than the correlation length of the system.  $\alpha_q$ is constant for a fractal surface and is equal to $\alpha$ defined in Equation 4, but is dependent on $q$ for a multi-affine surface. In Figure \ref{alpha-q} we have presented the roughness exponents obtained with this method for different orders $q$. Also, the time correlation function can be used to calculate $\beta$

\be
C(t)=\left[ \left< \left(h(x,t')-h(x,t+t') \right)^2 \right>_{x,t'} \right]^{1/2}\sim h^{\beta}
\ee

It can be seen that $\alpha$ is not constant for both experiments and simulations, with values in the range of $0.35<\alpha_q<0.43$, indicating that the growing surface is slightly multi-fractal. The other important fact is that these values are different from the value obtained with the first method $\alpha=0.53,\ 0.55$. This is also the case for $\beta$ (Table I). This means that maybe Family-Viscek scaling relations (Equation 5) do not apply to our system.  To check the correctness of the obtained roughness and growth exponents, we try to overlap the plots by scaling the height and roughness with Equation (5) for different system sizes $l$ (Fig 7). To do that, we first try to find $\alpha$ by plotting $\log(w/l^\alpha)$ for different system sizes $l$, vs. $\log(h)$ for different $\alpha$'s to see for which $\alpha$ the saturated tails of the curves match \cite{Family1, Family2}. We fit horizontal lines to the saturated part of each curve to find $w_s/l^\alpha$ for each $l$ and different values of $\alpha$. Then we calculate the standard deviation of them. The $\alpha$ which minimizes this standard deviation, is supposed to be the roughness exponent of the system (Figure 7(b) inset). We see that for $\alpha=0.29\pm0.01$ the standard deviation of the tails is minimum and the saturated part of different plots for different sizes have the best overlapping.

\begin{table}[ht!]
\small
\centering
\caption{The growth and roughness exponents calculated with different methods.}
\label{table1}
 \renewcommand{\arraystretch}{1.6}{
 \begin{tabular}{|l|l|l|l|l|}
\hline
method & work & \ \ $\beta$ & \ \ $\alpha$  \ \ &\ \ z \ \\
\hline
roughness & experiment &\  0.39 \ & \  0.53 \   & \ 1.4\ \\
\hline
roughness & simulation&\  0.32 \  &\  0.55 \  &\  1.7 \ \\
\hline
correlation & experiment&\   0.29 &\  0.40 \  &\ 1.4 \\
\hline
correlation &simulation&\  0.42 \  &\  0.36 \ &\  0.9\\
\hline
scaling &experiment&\ 0.55\  &\  0.29 \  &\  0.5\  \\
\hline
scaling &simulation&\ \  - &\  0.11 \ &\  \ - \\
\hline
 \end{tabular}}
\end{table}

\begin{figure}[h!]%[htbp] 
\centering
\centerline{\includegraphics[width=0.99 \columnwidth]{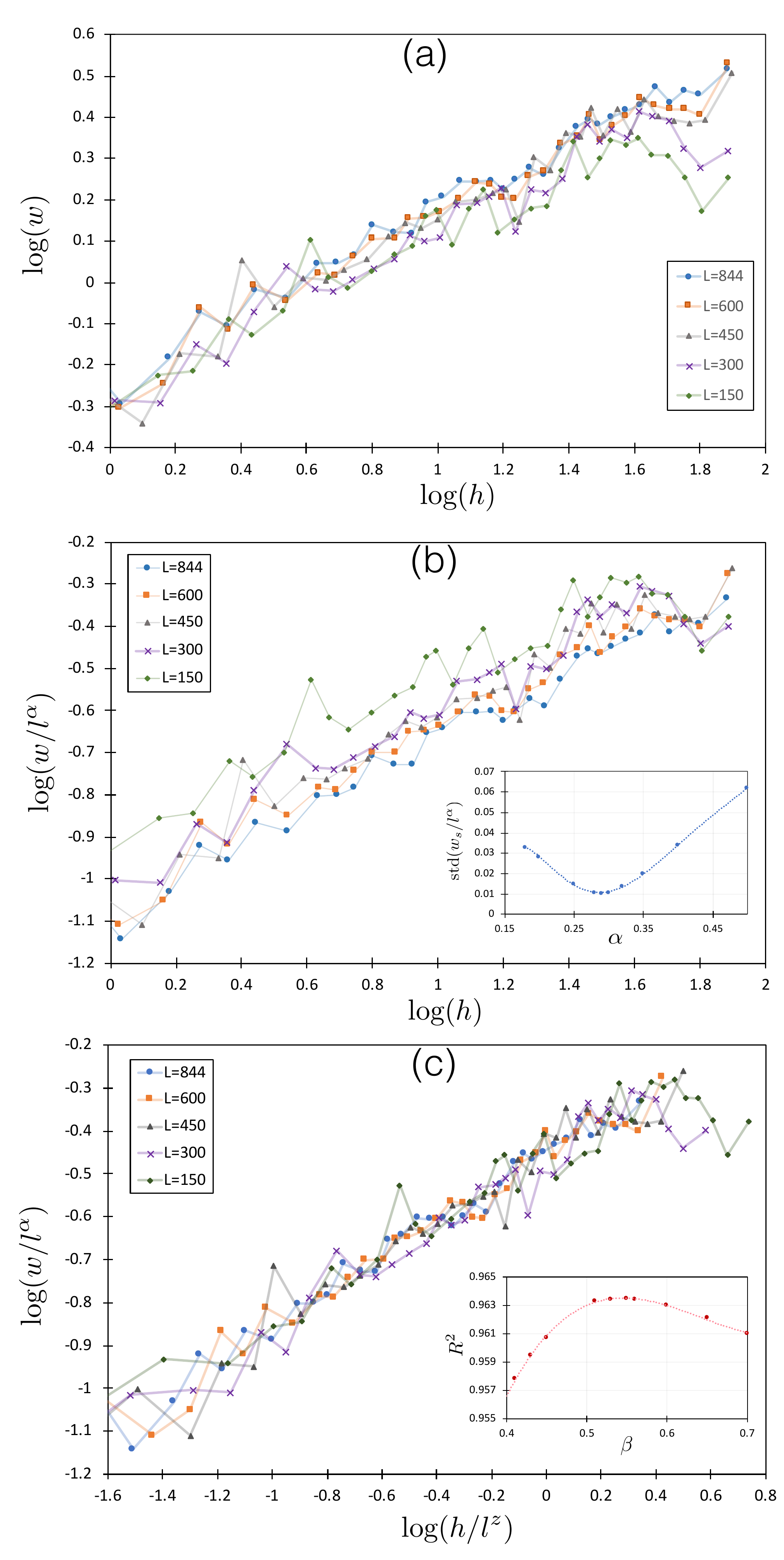}}
\caption{(a) The logarithm of the scaled width vs. the logarithm of the scaled mean height for different scaled system sizes $L$ for experimental data. (b) logarithm of the rescaled width as a function of $\log(h)$ for $\alpha=0.29$ corresponding to the best collapse of the saturated part of the curves, inset: standard deviation of the average saturated parts of different sizes as a function of $\alpha$. (c) log of the rescaled width vs. log of the rescaled height for $\beta=0.55$ corresponding to the best collapse of the first linear parts of the curves, inset: $R^2$ of the line fitting to the first part of curves of all system sizes vs. $\beta$.}
\label{scaling}
\end{figure}
 
 In the next step, we try to find $z=\alpha/\beta$ by finding the value for which the curves $\log(w/l^\alpha)$ vs. $\log(h/l^z)$ overlap \cite{Family1, Family2, Barabasi}. We change $z$ to find the value for which the first parts of the curves with power-law behavior collapse onto one curve. In order to do that, we plot all the data points for different system size curves together for the times before saturation. Then we fit one line to all the data and calculate the $R^2$ of the fit for different values of $z$ or equivalently $\beta=\alpha/z$ (Figure 7(c) inset). The best fit with the largest $R^2$ gives the growth exponent $\beta=0.55\pm0.2$. The simulation results did not collapse entirely on one curve, because the growth curve had two slopes (Figure 5(a)), and the first step gives a very small value for the roughness exponent $\alpha=0.11\pm0.1$. The values of the exponents driven using different methods are summarized in Table I.

\section{Discussion}

The first model proposed to describe the surface growth resulting from sediment granular matter was the continuous PDE equation by Edwards and Wilkinson \cite{EW}, which predicted a growth exponent of $\beta=1/4$ and a roughness exponent of $\alpha=1/2$ for 1+1 dimensions. Then Family \cite{Family2} developed a discrete lattice model for the same system, which added surface diffusion to a random deposition model. In this model, the deposit particles seek the local minimum of the height in a special range and settle there. This model gave growth and roughness exponents similar to EW model. For falling sticky particles, KPZ equation and the Ballistic Deposition discrete lattice model describe the dynamics of the growth front, giving the exponents $\beta=1/3$ and $\alpha=1/2$ \cite{KPZ, Ballistic}.

For real granular sedimenting systems with spherical or disc shape grains, the lattice model is not accurate, so the off-lattice models were proposed to study such systems. The first off-lattice simulation with disc-shape particles was done by Meakins and Jullien \cite{rain}. They allowed the sedimenting particle to roll over the particles beneath in two models: model II in which the particle rolls until it touches another particle or the base, and model III in which the particle rolls until it reaches the local minimum of the height and settles there. In contrast to Family's lattice model, both model gave a growth exponent $\beta$ close to the KPZ class (Table II, Rain model II $\&$ III). The same models was used by Csah{\'o}k and Vicsek \cite{Vicsek}, but they found different $\beta$'s for the two models, for the first model close to EW class, and for the second one close to KPZ (Table II, Csah{\'o}k I $\&$ II). 

The first experimental work on surface growth in a quasi-2-dimensional system of sedimenting particles was done by Kurnaz et al. \cite{Kurnaz1, Kurnaz2} using silica particles sedimenting in a viscous oil. The 1-dimensional growing front in their experiment had a steep hill shape, which resulted in high values for the roughness exponent. They observed two different roughness exponents for different length scales. We will only report the exponent for large length scales (larger than the particle size), because the first slope depends on the particle shape rather than the dynamics involved \cite{Reis}. Their results for the least viscous oil they used are reported in Table I (Kurnaz 96). In their next work, McCloud et al. used another setup with a moving funnel making a uniform front without a hill \cite{Kurnaz3}. They found that the roughness exponent is very sensitive to the particle deposition rate and increases with it. Their reported $\alpha$ can be found in Table I (Kurnaz 97). In a successive work \cite{Kurnaz4}, McCloud et al. studied the effect of the 3rd dimension on the roughness exponent, and saw no dependence (Table I, McCloud). In all the works with the second setup, they found roughness exponents less than 1/2 and they did not report the growth exponent. They also did a 3D simulation on their last experiment and found a different, single value roughness exponent (Table I, Cardak) \cite{Kurnaz5}. In all these experiments (except the first one with a hill), the reported exponents $\alpha$ are less than the theoretical prediction $\alpha=0.5$ both for EW and KPZ models, and $\beta$ is not reported. 

 \begin{table}[ht!]
\small
\centering
  \caption{The growth and roughness exponents reported in different references.}
  \label{table2}
 \renewcommand{\arraystretch}{1.6}{
 \begin{tabular}{|l|l|l|l|}
\hline
model & work & \ \ $\beta$ & \ \ $\alpha$ \\
\hline
Rain Model II \cite{rain}& simulation &  0.30 & \  -  \\
\hline
Rain Model III \cite{rain}& simulation &  0.31 & \  -  \\
\hline
Csah{\'o}k I \cite{Vicsek} & simulation&  0.31 & 0.44  \\
\hline
Csah{\'o}k II \cite{Vicsek} & simulation&  0.23 & 0.45  \\
\hline
Kurnaz 96 \cite{Kurnaz2} &experiment& 0.46 & 0.93  \\
\hline
Kurnaz 97 \cite{Kurnaz3} &experiment& - & 0.2-0.5  \\
\hline
McCloud \cite{Kurnaz4}  &experiment&  - & 0.31  \\
\hline
Cardak \cite{Kurnaz5}&simulation&  - & 0.40  \\
\hline
 \end{tabular}}
\end{table}

Our work is the first detailed study on the surface formed by deposition of sedimenting particles in a low-viscosity fluid, reporting both $\alpha$ and $\beta$ with different methods and also investigating the scaling and multi-fractality of the 1+1 surface formed. What we found with plotting the width of the interface as a function of height or local length, seemed to be close to KPZ class, both for experiments and simulations. In simulations, we had two different $\beta$'s for small and large scales, but we only consider the data for scales larger than a particle size. The simulation plot for $\alpha$ is a little strange, because it seems not to have a constant slope. We considered the average slope for this plot. Although, when we used correlation function method, we found different values for the growth and roughness exponents. For experimental data, we found a $\beta$ which was 0.1 less than the one derived with the first method, while for simulations, we found a $\beta$ which was 0.1 larger. Both data showed smaller $\alpha$ with this method. This resulted in a dynamic exponent $z<1$ for simulations.

In the next step, we investigated the multifractality of the surface, using the $q$'th order of $\alpha$. This showed that the system is multi-affine and thus cannot be described with any scaling theory. We tried to check if the system has Family-Vicsek scaling, by finding exponents that make all data collapse on one curve. For experimental data, the best collapse took place for very different values of $\alpha$ and $\beta$ corresponding to $z=0.5$. For simulations, the data did not collapse. All this measurements mean that our system, despite being very simple and expected to behave as EW or KPZ models, is a multi-affine system which does not obey the Family-Vicsek scaling. 

There are other theoretical predictions for systems with quenched disorder, correlated noise or power-low noise amplitude that give multifractality and $\alpha$'s larger than 0.5, but no theory gives $\alpha$ substantially less than one half \cite{KPZQ, power, correlated, correlated2}. Also, our system is unlikely to have quenched disorder or correlated noise. Thus, there should be other reasons behind the odd behavior of the system. Also, the hydrodynamic interactions have little effect on our particles, since our deposition rate is small and the fluid viscosity is low (despite the systems studied in \cite{Kurnaz3}). One major difference between our system and the other simple models for sedimenting grains is that our grains have dynamics throughout the whole experiment time. The grains are allowed to change their position after the first settlement. An oncoming grain can change the position of the grains with which it collides with, and those can also exert force on their neighboring grains and move them a bit \cite{impact}. So, the collision can be transmitted through the settled grains in a special range and change the position of other grains in that range. This restructuring going on through the whole process is the new feature of our simulation model and experiments. Besides, the particles at the bottom feel the pressure of the particles above, and this height-dependent pressure may also have an effect on the system dynamics.

\section{Conclusion}

In this work, we studied the surface growth resulting from the sedimentation of spherical polystyrene particles in ethanol in a vertical Hele-Shaw cell, which is a quasi-two dimensional system having a 1+1 dimensional growing interface. We performed experiments and MD simulations taking into account granular interactions throughout the experiment. We measured the growth and roughness exponents of the process using different methods, and derived different exponents. We showed that our system is multifractal and does not obey Family-Vicsek scaling, both for experiments and simulation. 

\section{Acknowledgement}

We thank S M Vaez, M D Niry, and A Saberi for helpful discussions and comments; R Shakoory for helping with  experiments; and L Bahmani and Z Zarei for their guidance on the simulations. 

{}
\end{document}